# Inverted Scanning Microwave Microscope for *In Vitro* Imaging and Characterization of Biological Cells

Marco Farina[1,*], Xin Jin[2], Gianluca Fabi[1], Eleonora Pavoni[1], Andrea di Donato[1], Davide Mencarelli[1], Antonio Morini[1], Francesco Piacenza[5], Richard Al Hadi[4], Yan Zhao[4], Tiziana Pietrangelo[3], Xuanhong Cheng[2], James C. M. Hwang[2]

This paper presents for the first time an innovative instrument called an inverted scanning microwave microscope (iSMM), which is capable of noninvasive and label-free imaging and characterization of intracellular structures of a live cell on the nanometer scale. In particular, the iSMM is sensitive to not only surface structures, but also electromagnetic properties up to one micrometer below the surface. Conveniently, the iSMM can be constructed through straightforward conversion of any scanning probe microscope, such as the atomic force microscope or the scanning tunneling microscope, with a simple metal probe to outperform traditional SMM in terms of ruggedness, bandwidth, sensitivity and dynamic range. By contrast, **the application of the traditional SMM to date has been limited to mainly surface physics and semiconductor technology, because the traditional SMM requires a fragile and expensive probe and is incompatible with saline solution or live biological cells.**

[1]Dip. di Ingegneria dell'Informazione, Università Politecnica delle Marche, Via Brecce Bianche, 60131 Ancona, Italy

[2]Dept. of Electrical and Computer Engineering, Lehigh University, Bethlehem, PA 18015 USA

[3]Dept. of Neuroscience, Imaging and Clinical Sciences, University G d'Annunzio Chieti-Pescara, I-66013 Chieti, Italy

[4]Alcatera LLC, UCLA Anderson Accelerator, Los Angeles, CA, USA

[5]Advanced Technology Center for Aging Research, Scientific Technological Area, IRCCS INRCA, Via Birarelli 8, 60121 Ancona, Italy

*Corresponding author. E-mail: m.farina@univpm.it

In a scanning probe microscope such as the atomic force microscope (AFM) or the scanning tunneling microscope (STM), a probe in the form of a sharp stylus is scanned across the sample within 1 nm of its surface, while variation of the interaction between the probe and the sample is recorded *(1)*. For example, variations of force and current are recorded in the AFM and STM, respectively. The interaction can also be through an evanescent electromagnetic field, such as that in the near-field scanning optical microscope (SNOM) and the scanning microwave microscope (SMM) *(2)*. In this case, the spatial resolution is determined by the sharpness of the probe tip rather than the electromagnetic wavelength. Despite the much longer wavelength of the microwave than that of the optical wave, the SMM has the following advantages over the SNOM: (i) noninvasiveness because the energy of microwave photons is on the order 10 µeV, (ii) sensitivity to optically opaque materials *(3)*, (iii) spectroscopy over many decades of frequencies from the same microwave source, and (iv) sensitivity to dielectric permittivity of frequencies relevant to most electronic and biological functions. For these advantages, the SMM can be explored for many other applications beside the current focus on surface physics and semiconductor technology *(4)*.

In principle, the most impactful application of the SMM is in noninvasive, label-free imaging and characterization of live cells, organelles, bacteria and viruses. However, so far there have only been a few reports on the application of SMM in biology, and they are limited to dead or barely surviving samples *(5, 6)*. This is mainly because the microwave probe is often incompatible with the saline solution and, even if it survives the saline solution, it is rendered insensitive by the parasitic interaction between the probe body and the surrounding ground (Fig. 1A) *(7, 8)*. Because a conducting sample holder is usually used to maximize the reflection of the microwave signal, the parasitic interaction can be orders-of-magnitude larger than the intrinsic interaction between the probe tip and the sample. It is even worse when the probe is immersed in the saline solution, because the solution has much higher dielectric permittivity than that of air. Currently, to boost the sensitivity of an SMM amidst the parasitic interaction, a resonance circuit is often used *(3)*, which precludes broadband spectroscopy. On the other hand, broadband SMM is desirable because (i) it provides information at relevant frequencies, (ii) it allows time-gated filtering of unwanted signals through post-measurement data processing *(5)*, and (iii) it enables microwave tomography.

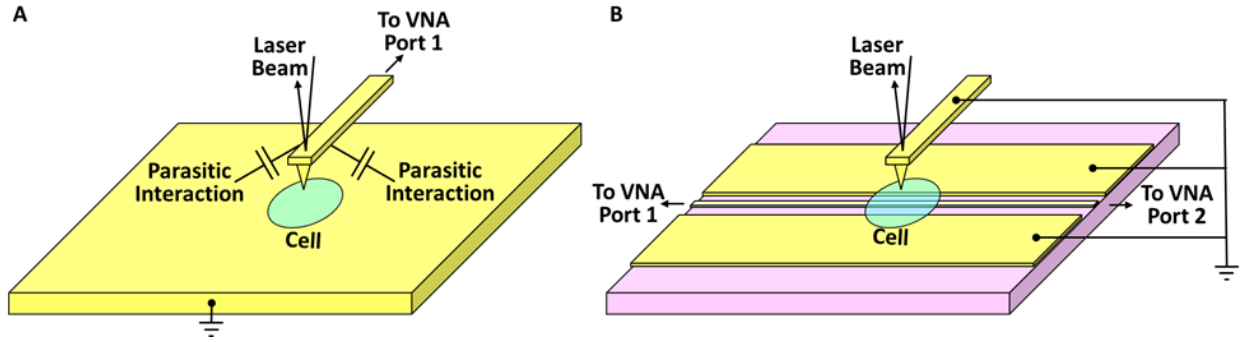

**Fig. 1**. Schematics of (A) a traditional AFM-modified SMM and (B) an inverted SMM. In (A), one-port microwave measurement is performed through the AFM probe, which suffers from parasitic interaction between the probe body and the surrounding ground. The parasitic interaction is aggravated when the probe is immersed in a conducting solution. In (B), two-port microwave measurement is performed through the input and output ports of a coplanar waveguide as part of the sample holder. The parasitic interaction between the probe body and the surround is minimized because the probe is grounded, too.

Typically, an SMM is modified from an AFM or an STM, so that the probe can be scanned at a fixed height above the sample through the feedback of force or current. In either case, the microwave signal is injected through the probe by a vector network analyzer (VNA), and the reflected signal is also sensed by the VNA via the probe. The ratio of the reflected and the injected signals, the reflection coefficient, can be used to determine the spreading impedance or dielectric permittivity of the sample, after proper calibration and analysis. Such a one-port reflection measurement usually has a dynamic range of 40-60 dB as limited by directional couplers.

To minimize the parasitic interaction and to boost the SMM sensitivity without resorting to a resonance circuit, we propose a novel technique based on an inverted SMM (iSMM). As shown schematically in Fig. 1B, in an iSMM the scanning probe is always grounded and the microwave signal is injected through a transmission line (e.g., coplanar waveguide, slot line) as part of the sample holder. Unlike the traditional SMM probe, the transmission line can have broadband impedance match over many decades of frequency *(9)*. The input and output of the transmission line are connected to the VNA, so that both reflection and transmission coefficients are measured. Such a two-port measurement usually has a dynamic range of 120–140 dB, which makes it easier to sense the tiny perturbance caused by the grounded probe when it scans across the sample.

According to the reciprocity theory of electromagnetics, the intrinsic interaction between the probe tip and the sample is the same whether the microwave signal is injected through the probe or the sample. However, with the microwave signal injected through the sample and the probe grounded, the parasitic interaction between the probe body and the surround is greatly reduced, because most of the surround is grounded in any case. Thus, compared to a traditional SMM, an iSMM can have wider dynamic range, higher sensitivity, and broader bandwidth. Additionally, the probe can be a simple, rugged and bio-compatible metal stylus. Meanwhile, whether the iSMM is modified from an AFM or STM, the original AFM or STM function is intact so that an iSMM image can be obtained simultaneously with an AFM or STM image.

To demonstrate the new technique, we used an AFM-modified iSMM on Jurkat (dried only) and L6 cells (both live and dried). The detailed procedure is described in Supplementary Materials. Fig. 2 compares the AFM and iSMM images of dried Jurkat cells. It can be seen that the quality of the iSMM image is at least as good as that of the AFM image.

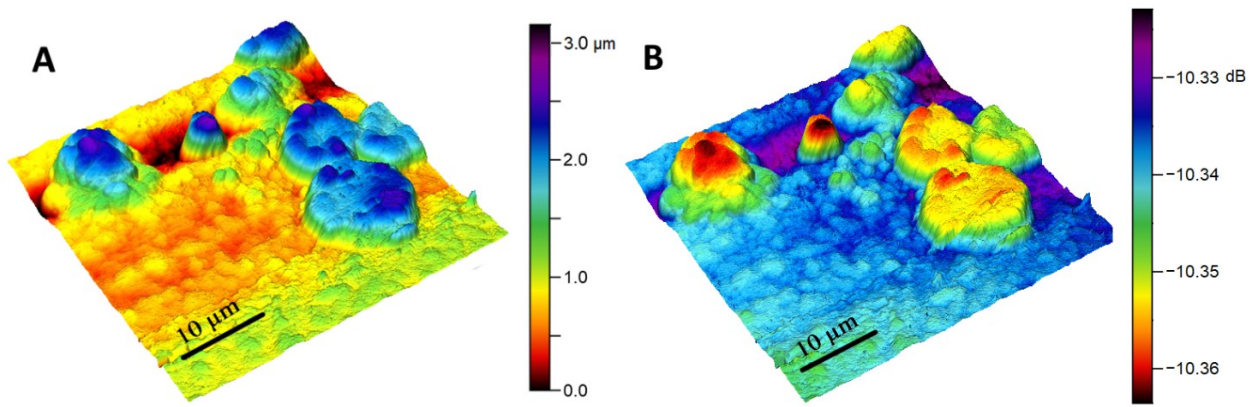

**Fig. 2. Simultaneous (A) AFM and (B) iSMM images of dried Jurkat cells. The iSMM image is based on the magnitude of the reflection coefficient at 4 GHz.**

Fig. 3 compares the AFM and iSMM images of a live L6 cell in saline solution. The difference between AFM and iSMM is that iSMM is sensitive to the properties of intracellular structures below the surface. After proper calibration, the iSMM can quantify the dielectric permittivity of intracellular structures. To the best of our knowledge, this is the first successful image of a live cell in physiological condition, whether by SMM or iSMM. This is also among the best-quality images formed by the transmission coefficient measured by a two-port SMM or iSMM *(10, 11)*.

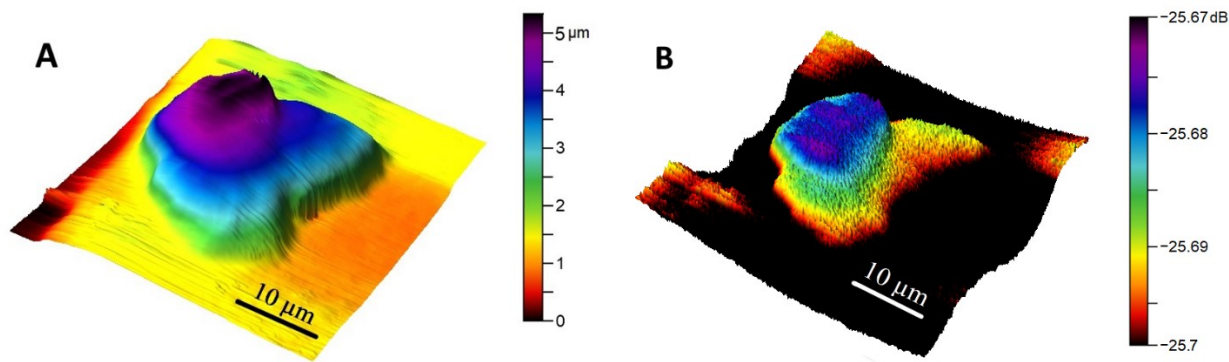

**Fig. 3. Simultaneous (A) AFM and (B) iSMM images of a live L6 cell in saline solution. The iSMM image is based on the magnitude of the transmission coefficient at 3.4 GHz.**

The calibration of SMM is not trivial and SMM has been used mostly for imaging instead of quantitative characterization. This is because most SMM models consider only the intrinsic interaction between the probe tip and the sample, ignoring the afore-mentioned parasitic interaction between the probe body and the surround. Using an innovative calibration procedure *(12)* detailed in Supplementary Materials, Fig. 4 illustrates the effect of calibration on the iSMM image of a dried L6 cell. Fig. 4A is the AFM topography image. Fig. 4B is the iSMM capacitance image corrupted by topography. Fig. 4C is the iSMM dielectric constant image with the topography effect removed. As expected, the dried cell exhibited ridges near its periphery, but rather uniform dielectric constant of 2.8 ± 0.7 across the cell. This value is comparable to that of lipid bilayers in electrolyte solution *(13)*, but is lower than that of dried *E. coli* bacteria *(14)*.

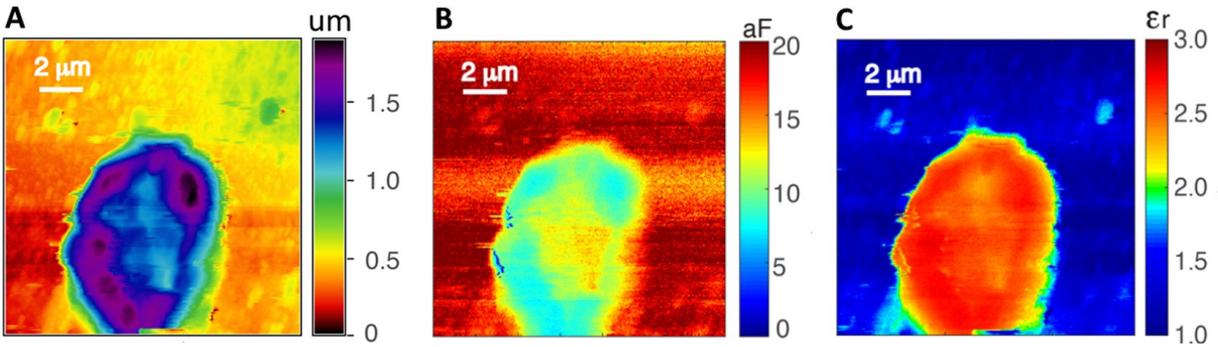

**Fig. 4. (A) AFM topography, (B) iSMM capacitance, and (C) iSMM dielectric constant images of a dried L6 cell. The iSMM image is based on the magnitude of the transmission coefficient at 6.2 GHz.**

In conclusion, we have demonstrated a novel iSMM for quantitative imaging and characterization on the nanometer scale, which can be applied through a straightforward modification of any scanning probe microscope. The technique is broadband, highly sensitive, label-free, and non-invasive yet sensitive to subsurface electromagnetic properties. With a simple and rugged metal probe that is always grounded, it can be easily made biocompatible. Using the iSMM, a live cell in saline solution was successfully imaged which has not been possible with a traditional SMM. Thus, the iSMM should broaden the application of SMM to many applications beyond the current focus on surface physics and semiconductor technology. With the broadband capacity of iSMM, work is in progress for microwave tomography, the equivalent of optical coherent tomography but with far less invasive microwave photons.

**ACKNOWLEDGMENTS**

**Funding:** This work was funded in part by the US Army under Grants W911NF-14-1-0665, W911NF-17-1-0090 and W911NF-17-P-0073, and the US Air Force under Grants bacterial cells at gigahertz frequencies by scanning microwave microscopy," *ACS Nano* **10**, 280–288 (2016).

**SUPPLEMENTARY MATERIALS**

Supplementary Text

Figs. S1 to S7

Reference (15, 16)